\documentclass[final,twocolumn,5p]{elsarticle}

\usepackage{lineno,hyperref}
\usepackage{amsmath}
\usepackage{amssymb}
\usepackage{braket}
\usepackage{multirow}
\usepackage{color}
\biboptions{numbers,sort&compress}
\usepackage{lipsum}
\usepackage{longtable}

\modulolinenumbers[5]

\begin{document}

\begin{frontmatter}

\title{Hyperfine components of all rovibrational quadrupole transitions \\ in the H$_{2}$ and D$_{2}$ molecules}


\author[mymainaddress]{Hubert J\'{o}\'{z}wiak\corref{mycorrespondingauthor}}
\ead{hubert.jozwiak@stud.umk.pl/hubj96@gmail.com}
\cortext[mycorrespondingauthor]{Corresponding author}

\author[mysecondaryaddress]{Hubert Cybulski}

\author[mymainaddress]{Piotr Wcis{\l}o}
\ead{piotr.wcislo@umk.pl}

\address[mymainaddress]{Institute of Physics, Faculty of Physics, Astronomy and Informatics, Nicolaus Copernicus University in Torun, Grudziadzka 5, 87-100 Torun, Poland}
\address[mysecondaryaddress]{Institute of Physics, Kazimierz Wielki University, ul.
Powsta{\'n}c{\'o}w Wielkopolskich 2, 85-090 Bydgoszcz, Poland}

\begin{abstract}
We report results of a theoretical investigation of hyperfine interactions in two homonuclear isotopologues of the hydrogen molecule: H$_{2}$ and D$_{2}$. We present a set of hyperfine coupling constants: spin-rotation, spin-spin dipole and, in the case of the D$_{2}$ molecule, electric quadrupole coupling constants for all bound states of the two isotopologues in their ground electronic $X^{1}\Sigma^{+}_{g}$ state. We provide a list of positions and intensities of 220~997~hyperfine components of 16~079 rovibrational quadrupole transitions of the O, Q and S branches. The positions and intensities of the hyperfine components are necessary for a reliable interpretation of accurate measurements of rovibrational transition frequencies in H$_{2}$ and D$_{2}$, which are used for tests of the quantum electrodynamics of molecules and searches for new physics beyond the Standard Model.
\end{abstract}


\end{frontmatter}


\section{Introduction}
Theoretical studies of the rovibrational structure of the ground electronic state in molecular hydrogen achieve remarkable accuracy in the determined dissociation energy~\cite{Puchalski2016,Puchalski2017,Puchalski2018,Puchalski2019,Puchalski2019b} and rovibrational energy levels \cite{Pachucki2014,Pachucki2015,Pachucki2016,Pachucki2018,Pachucki2018b,Czachorowski2018,Komasa2019}. The progress in the theoretical investigations is enhanced by recent experiments, which aim at the sub-MHz accuracy of both the dissociation energy \cite{Altmann2018,Holsch2019} and the rovibrational transition frequencies in the ground electronic state \cite{Dickenson2013,Mondelain2016,Cozijn2018,Tao2018,Fasci2018,Wcislo2018,Lai2019,Beyer2019,Zaborowski2020,Wojtewicz2020,Mondelain2020,Diouf2020,Fast2020}. Comparison between theoretical calculations and accurate measurements can be used as a test for the quantum electrodynamics for molecules~\cite{Komasa2019}, determination of the proton-charge radius~\cite{Puchalski2016,Pachucki2016,Altmann2018,Holsch2019} or even searches for physics beyond Standard Model~\cite{Ubach2016}, i.e. by putting constraints on extra dimensions~\cite{Salumbides2013} or new forces~\cite{Salumbides2015}. 

As dipole transitions in the homonuclear isotopologues of hydrogen molecule (H$_{2}$, D$_{2}$ and T$_{2}$) are forbidden, the experimental studies are limited to quadrupole transitions, which are much weaker than dipole transitions in the HD molecule~\cite{Kassi2011,Kassi2012,Campargue2012,HITRAN}. The quadrupole transitions in H$_{2}$ and D$_{2}$ have never been saturated and the most accurate experiments were performed using Doppler-limited cavity-enhanced techniques. Nevertheless, the best measurements for the D$_{2}$ molecule, performed for the S(2) line from the first overtone, reached the accuracies of 500~kHz~\cite{Mondelain2016}, 400~kHz~\cite{Wcislo2018} and even 161~kHz~\cite{Zaborowski2020}. The first overtone in D$_{2}$ was also recently studied by W\'{o}jtewicz~\textit{et~al.}~\cite{Wojtewicz2020}, who reported 870 and 999~kHz combined uncertainties for the positions of the S(3) and S(4) lines, respectively. Mondelain~\textit{et~al.}~\cite{Mondelain2020}, estimated the uncertainty of the frequencies of the S(0) and Q(1)-Q(4) D$_{2}$ transitions between 0.3 and 1.83~MHz. In the case of the H$_{2}$ molecule, the most accurate measurements that probed rovibrational levels directly, reached the accuracy of 6.6~MHz~\cite{Wcislo2016}, while measurements based on subtraction of the energies of two electronic transitions achieved the accuracy of 4.5~MHz~\cite{Dickenson2013,Niu2014} for ${\nu = 0, N=0}$ and ${\nu=1, N =0}$ states and 1.5~MHz~\cite{Beyer2019} for the ${\nu = 0, N=0}$ and ${\nu=0, N =1}$ states.

Such accurate measurements of rovibrational transitions require including in the spectral analysis the hyperfine structure originating from the interactions involving nuclear spins and (in the case of deuteron-bearing isotopologues) electric quadrupole moment of the nuclei. The disagreement between recent measurements of rovibrational dipole transitions in the HD molecule \cite{Cozijn2018,Tao2018,Fasci2018} has been attributed to the underlying hyperfine structure of rovibrational states, which were recently studied by Diouf~\textit{et~al.}~\cite{Diouf2019}. Theoretical studies of the hyperfine components of rovibrational dipole lines in HD were reported~\cite{Dupre2020,Jozwiak2020,Komasa2020} but no such a study concerning quadrupole transitions in H$_{2}$ or D$_{2}$ was performed.

Here, we provide a comprehensive list of all 220~997 hyperfine components of 16~079 rovibrational quadrupole transitions in H$_{2}$ and D$_{2}$. We report the list of rovibrationally-averaged hyperfine coupling constants, i.e. spin-rotation, spin-spin dipole, and, in the case of D$_{2}$ molecule, electric quadrupole constant, for all bound states of both molecules.

The paper is organized as follows: Section~\ref{Sec:HyperfineTheory} provides a discussion of the components of the hyperfine Hamiltonian for both molecules. In Sec.~\ref{Sec:CouplingConstants}, we discuss the hyperfine coupling constants, which are calculated for all bound states of H$_{2}$ and D$_{2}$, and in Sec.~\ref{Sec:EnergyLevels} we discuss the hyperfine splittings of the rovibrational levels. Sections~\ref{Sec:LineIntensities} and ~\ref{Sec:HFLineIntensities} give descriptions of the line intensities of the rovibrational and hyperfine transitions. In Sec.~\ref{Sec:example}, we present examples of calculated positions and intensities for two quadrupole lines in H$_{2}$ and D$_{2}$. In Sec.~\ref{Sec:Conclusion}, we conclude our results.

\section{Hyperfine interactions}
\label{Sec:HyperfineTheory}
In this section, we provide an analysis of the hyperfine (HF) structure Hamiltonian, $\mathcal{H^{\rm HF}}$, of the H$_{2}$ and D$_{2}$ molecules. In the case of the H$_{2}$ molecule, the hyperfine structure of rovibrational levels originates from the interactions involving nuclear spin of the proton ($I_{\rm H} = \frac{1}{2}$). As the H$_{2}$ molecule consists of two protons, the two nuclear spins couple and the total nuclear spin, $I$, might be either 0, for \textit{para}-H$_{2}$, or 1, for \textit{ortho}-H$_{2}$. Due to the symmetry relations imposed on fermions, \textit{para}-H$_{2}$ and \textit{ortho}-H$_{2}$ exhibit rotational structures which involve either even or odd rotational quantum numbers, $N$, only. The rovibrational transitions in \textit{para}-H$_{2}$ do not exhibit any hyperfine structure, due to the fact that $I$ equals 0. Hence, we confine the analysis to \textit{ortho}-H$_{2}$ and we consider the nuclear spin-rotation and spin-spin dipole coupling terms in the hyperfine Hamiltonian.

On the other hand, since the deuteron is a particle of spin $I_{\rm D} = 1$, the D$_{2}$ molecule obeys the symmetry relations for bosons. The coupling of the two nuclear spins leads to three possible values of the total spin, $I = 0, 1$ and $2$. \textit{para}-D$_{2}$ corresponds to the $I=0$ and $2$ cases, and their rotational structure consists of states described with odd rotational quantum numbers, while the rotational structure of \textit{ortho}-D$_{2}$ ($I = 1$) involves only even rotational states, $N$. As the nuclear spin of the deuteron is greater than~$\frac{1}{2}$, the interaction of the electric quadrupole moment with the molecular electric field gradient (EFG) must be taken into account, on top of the spin-rotation and spin-spin dipole coupling.

In both cases, we use a coupled basis to calculate matrix elements of the hyperfine Hamiltonian. The coupled basis is formed from coupling of two nuclear spins $I_{\rm X}$ ($\rm X = \rm H$ or $\rm D$) to form the total nuclear spin $I$, which is further coupled to the rotational angular momentum $N$, to form the total angular momentum, $F$. The state vector in this representation is given as $\ket{\nu; N I F m_{\rm F}}$, where $\nu$ and $m_{\rm F}$ are the vibrational quantum number and the projection of the total angular momentum on the laboratory-fixed frame, respectively. We restrict the analysis to the ground electronic $X^{1}\Sigma^{+}_{g}$ state and we neglect any coupling between vibrational states.

\subsection{H$_{2}$ molecule}
We consider the effective hyperfine Hamiltonian, $\mathcal{H}^{\rm HF}_{\rm H_{2}}$, which consists of two terms \cite{Code1971,BrownCarrington2003}:
\begin{equation}
    \label{eq:hyperfineHamiltonian}
    \mathcal{H}^{\rm HF}_{{\rm H_{2}}} = \mathcal{H}_{\rm {nsr}} +\mathcal{H}_{\rm {dip}}.
\end{equation}
The former term, $\mathcal{H}_{\rm {nsr}}$, corresponds to the nuclear spin-rotation coupling, while the latter, $\mathcal{H}_{\rm dip}$, denotes the dipolar interaction between the two nuclear spins.

The nuclear spin-rotation coupling originates from the interaction of a nuclear magnetic moment with the magnetic field resulting from the overall rotation of the molecule. As we consider the diatomic molecule, the strength of the coupling is given by a simple scalar value, namely the spin-rotation constant~\cite{Ramsey1952a,White1955,Reid1974,Flygare1964,BrownCarrington2003}, which consists of the sum of the nuclear and electronic contributions to the molecule's magnetic field, averaged over the rovibrational wavefunction of the molecule. This part of the Hamiltonian is given as:
\begin{equation}
\label{eq:nsrH2}
    \mathcal{H}_{\rm {nsr}} = c_{\rm nsr} \mathbf{I}\cdot \mathbf{N},
\end{equation}
where $c_{\rm nsr}$ denotes the nuclear spin-rotation coupling constant of the H$_{2}$ molecule. 

The matrix elements of the spin-rotation interaction can be evaluated using the spherical tensorial algebra of Fano and Racah~\cite{Fano1959, BrownCarrington2003}. The total nuclear spin, $\mathbf{I}$, and the rotational angular momentum, $\mathbf{N}$, can be represented as rank-1 spherical tensors. This part of the Hamiltonian is diagonal and the evaluation of the matrix elements is straightforward:
\begin{align}
\begin{split}
    \label{eq:nsr-matrixelements}
    &\braket{\nu;N' I'  F' m_{\rm F}'|\mathcal{H}_{\rm {nsr}}|\nu;N I F m_{\rm F}} = \delta_{NN'} \delta_{II'}\delta_{FF'}\delta_{m_{\rm F}m_{\rm F}'}\\
    &\times \frac{c_{\rm nsr}}{2} \Big(F\left(F+1\right)-I\left(I+1\right)-N\left(N+1\right)\Big).
\end{split}
\end{align}

The second term in the Hamiltonian from Eq.~\eqref{eq:hyperfineHamiltonian} corresponds to the magnetic dipole interaction between the nuclear magnetic moments of the protons. It is given as:

\begin{align}
\begin{split}
    \label{eq:dipdip}
        \mathcal{H}_{\rm dip} = g_{\rm I}^{2}\mu_{\rm N}^{2}\frac{\mu_{0}}{4\pi} \left[\frac{\mathbf{I}_{\rm 1}\cdot \mathbf{I}_{\rm 2}}{R^{3}}-\frac{3(\mathbf{I}_{\rm 1}\cdot \mathbf{R})(\mathbf{I}_{\rm 2}\cdot \mathbf{R})}{R^{5}}\right],
\end{split}
\end{align}
where $g_{\rm I}$ is the $g$ factor of the nucleus (here: proton), $\mu_{\rm N}$ is the nuclear magneton, $\mu_{0}$ is the vacuum permeability, and $R$ is the internuclear distance.

In order to employ the spherical tensorial algebra, we represent the nuclear angular momenta $\mathbf{I}_{\rm 1}$ and $\mathbf{I}_{\rm 2}$ as spherical tensors of rank 1 and couple them to form the rank-2 tensor $\mathrm{T}^{(2)}(\mathbf{I}_{1}, \mathbf{I}_{2})$. The resulting spherical tensor is coupled with $\mathrm{T}^{(2)}(\mathbf{C})$, the rank-2 tensor describing the spherical harmonics $Y_{2q}$ associated with the transformation of the rovibronic wavefunction from the laboratory-fixed to the molecule-fixed frame of reference. The final form of the spin-spin dipole interaction Hamiltonian, $\mathcal{H}_{\rm dip}$, is given as~\cite{BrownCarrington2003}:

\begin{align}
\begin{split}
    \label{eq:dipdip2}
        \mathcal{H}_{\rm dip} = g_{I}^{2} \mu_{\rm N}^{2}\frac{\mu_{0}}{4\pi} \sqrt{6} \mathrm{T}^{(2)}(\mathbf{C})\cdot \mathrm{T}^{(2)}(\mathbf{I}_{\rm 1}, \mathbf{I}_{\rm 2}),
\end{split}
\end{align}
and the matrix elements in the coupled basis are:
\begin{align}
\begin{split}
    \label{eq:dipdip-matrixelements}
    &\braket{\nu;N'I'F' m_{\rm F}'|\mathcal{H}_{\rm {dip}}|\nu;N I F m_{\rm F}} \\
    &=\delta_{FF'}\delta_{m_{\rm F}m_{\rm F}'}  (-1)^{N+I+F+N'} \sqrt{30} c_{\rm dip}\\
&\times \sqrt{(2I+1)(2I'+1)(2N+1)(2N'+1)}\\
    &\times
    \sqrt{I_{\rm 1}(I_{\rm 1}+1)(2I_{\rm 1}+1)I_{\rm 2}(I_{\rm 2}+1)(2I_{\rm 2}+1)}\\
&\times
\begin{pmatrix}
    N' & 2 & N \\
    0 & 0 & 0
\end{pmatrix} 
\begin{Bmatrix}
I&N&F\\
N'&I'&2
\end{Bmatrix} 
\begin{Bmatrix}
  I_{\rm 1} & I_{\rm 1} & 1 \\
  I_{\rm 2} & I_{\rm 2} & 1\\
  I & I' & 2 
\end{Bmatrix},
\end{split}
\end{align}
where we introduced the coupling constant:
\begin{equation}
    \label{eq:dipdip-couplingconstant}
    c_{\rm dip} = g_{\rm H}\mu_{\rm N}^{2}\frac{\mu_{0}}{4\pi} \braket{\nu N|\frac{1}{R^{3}}|\nu N},
\end{equation}
which involves the expectation value of the $1/R^{3}$ term in given rovibrational state. The quantity in parenthesis is the Wigner 3-\textit{j} symbol and the quantities in the small and large curly brackets are the Wigner 6-\textit{j} and 9-\textit{j} symbols, respectively. In the case of the \textit{ortho}-H$_{2}$ molecule ($I_{1} = I_{2} = \frac{1}{2}$ and $I=I'=1$)  Eq.~\eqref{eq:dipdip-matrixelements} reduces to 
\begin{align}
\begin{split}
    \label{eq:dipdip-matrixelements-H2}
    &\braket{\nu;N'I(=1) F' m_{\rm F}'|\mathcal{H}_{\rm {dip}}|\nu;N I(=1) F m_{\rm F}} \\
    &=-\delta_{FF'}\delta_{m_{\rm F}m_{\rm F}'}  (-1)^{N+F+N'} \sqrt{\frac{15}{2}} c_{\rm dip}\\
&\times \sqrt{(2N+1)(2N'+1)}
\begin{pmatrix}
    N' & 2 & N \\
    0 & 0 & 0
\end{pmatrix} 
\begin{Bmatrix}
1&N&F\\
N'&1&2
\end{Bmatrix} .
\end{split}
\end{align}

\subsection{D$_{2}$ molecule} 
\label{Sec:HyperfineD2}
In the case of the D$_{2}$ molecule, the effective Hamiltonian involves three terms:
\begin{equation}
    \label{eq:hyperfineHamiltonianD2}
    \mathcal{H}^{\rm HF}_{\rm D_{2}} = \mathcal{H}_{\rm {nsr}} +\mathcal{H}_{\rm {dip}}+\mathcal{H}_{\rm quad}.
\end{equation}
The first two terms as well as the corresponding matrix elements are essentially the same as in the previous section  (with the exception of the $g$-factor and coupling constants). Note that in the case of \textit{para}-D$_{2}$, a possible coupling between the $I = 0$ and $I = 2$ states may arise. 

Apart from the spin-rotation and spin-spin dipole coupling, one has to take into account the interaction of the electric quadrupole moment of the deuterium nuclei with the molecular electric field gradient. This term can be represented using the spherical tensorial algebra as:
\begin{align}
\begin{split}
    \label{eq:quad}
    \mathcal{H}_{\rm quad} = -e \sum_{k} \mathrm{T}^{(2)}(\mathbf{Q}_{k})\cdot \mathrm{T}^{(2)}(\mathbf{\nabla \mathbf{E}_{k}}),
\end{split}
\end{align}
where the sum runs over the two deuteron nuclei, labeled with~$k$. The quadrupole moment as well as EFG are represented as spherical tensors of rank 2 (see Chapter 8.4 of Ref.~\cite{BrownCarrington2003} for more details). Matrix elements of the quadrupole interaction are given as:
\begin{align}
\begin{split}
    \label{eq:quad-matrixelements}
    &\braket{\nu;N' I' F' m_{\rm F}'|\mathcal{H}_{\rm {quad}}|\nu;N I F m_{\rm F}} \\
    &=\delta_{FF'}\delta_{m_{\rm F}m_{\rm F}'} (-1)^{N+I'+F+I_{1}+I_{2}+N'} \sqrt{(2N+1)(2N'+1)}\\
&\times \sqrt{(2I+1)(2I'+1)}
\begin{pmatrix}
        N' & 2 & N \\
        0 & 0 & 0
    \end{pmatrix}
\begin{Bmatrix}
I' & N' & F\\
N  & I  & 2
\end{Bmatrix} \\
&\times \Biggl[
(-1)^{I}
\frac{c_{\rm Q_{1}}}{4}
\begin{Bmatrix}
I_{1} & I' & I_{2}\\
I  & I_{1}  & 2
\end{Bmatrix}
\begin{pmatrix}
          I_{\rm 1} & 2 & I_{\rm 1} \\
          -{I_{\rm 1}} & 0 & {I_{\rm 1}}
\end{pmatrix}^{-1} 
     \\   &+(-1)^{I'}
     \frac{c_{\rm Q_{2}}}{4}
\begin{Bmatrix}
I_{2} & I' & I_{1}\\
I  & I_{2}  & 2
\end{Bmatrix}
\begin{pmatrix}
          I_{\rm 2} & 2 & I_{\rm 2} \\
          -{I_{\rm 2}} & 0 & {I_{\rm 2}}
     \end{pmatrix}^{-1}\Biggr],
\end{split}
\end{align}
where the quadrupole coupling constant:
\begin{equation}
    \label{eq:quadconstant}
    c_{\rm Q_{k}} = eQ_{\rm D_{k}}q_{0_{k}},
\end{equation}
involves the electric quadrupole moment of the $k$-th nucleus, $Q_{\rm D_{k}}$, defined as the expectation value of the $Q_{33}$ element of the traceless and symmetric nuclear quadrupole moment tensor $Q_{ij}$, in the spin-stretched state:
\begin{equation}
    \label{eq:quadmoment}
    Q_{\rm D_{k}} = \braket{I_{k},m_{I_{k}}(=I_{k})|Q_{33}|I_{k},m_{I_{k}}(=I_{k})},
\end{equation}
and the rovibrationally-averaged EFG at the position of the $k$-th nucleus, $q_{0_{k}}$, which is the expectation value of the $V_{33}$ component of EFG tensor, $V_{ij}$, in a given rovibrational state. 
In the case of a homonuclear molecule such as D$_{2}$, where $I_{1}=I_{2}$, Eq.~\eqref{eq:quad-matrixelements} simplifies to:

\begin{align}
\begin{split}
    \label{eq:quad-matrixelementsHomo}
    &\braket{\nu;N' I' F' m_{\rm F}'|\mathcal{H}_{\rm {quad}}|\nu;N I F m_{\rm F}} \\
    &=\delta_{FF'}\delta_{m_{\rm F}m_{\rm F}'} (-1)^{N+I'+F+N'}\frac{c_{\rm Q}}{4}\left[(-1)^{I}+(-1)^{I'}\right]\\
    &\times  \sqrt{(2N+1)(2N'+1)(2I+1)(2I'+1)} 
    \begin{pmatrix}
        N' & 2 & N \\
        0 & 0 & 0
    \end{pmatrix} \\
    &\times 
     \begin{pmatrix}
          I_{\rm 1} & 2 & I_{\rm 1} \\
          -{I_{\rm 1}} & 0 & {I_{\rm 1}}
     \end{pmatrix}^{-1}
\begin{Bmatrix}
I' & N' & F\\
N  & I  & 2
\end{Bmatrix}
\begin{Bmatrix}
I_{1} & I' & I_{1}\\
I  & I_{1}  & 2
\end{Bmatrix}.
\end{split}
\end{align}
Similarly to the spin-spin interaction, the quadrupole coupling mixes the $N$ and $N'=N\pm2$ states as well as (in the case of \textit{ortho}-D$_{2}$) the $I$ and $I'=I\pm2$ states. 

\section{Hyperfine coupling constants}
\label{Sec:CouplingConstants}
Following our previous work~\cite{Jozwiak2020}, we determine the dependence of the coupling constants on the internuclear distance, $R$, and we calculate their expectation values in a given rovibrational state. Due to very large datasets for both molecules, the coupling constants are given in the Supplementary Materials~\cite{SupMat}. Similarly as we did in the case of the HD molecule~\cite{Jozwiak2020}, we do not include the $N=0$ levels, since none of the analyzed hyperfine terms from Eqs.~\eqref{eq:hyperfineHamiltonian} and \eqref{eq:hyperfineHamiltonianD2} affect the ground rotational levels.

The rovibrational wavefunctions of the H$_{2}$ and D$_{2}$ molecules, $\chi_{\nu, N} (R) = \braket{R|\nu N}$, are numerical solutions of the Schr\"{o}dinger equation for the isolated molecules within the Born-Oppenheimer approximation. We made use of the potential energy curve of Schwenke~\cite{Schwenke1988} and solved the Schr\"{o}dinger equations using the discrete variable representation (DVR) method. We obtained 146 and 596 wavefunctions for the \textit{ortho}-H$_{2}$ and D$_{2}$ molecules, respectively. Our set of wavefunctions does not include a very weakly bound state ${\nu=21, N=1}$ in D$_{2}$. The corresponding dissociation energy of this state is 0.0491~cm$^{-1}$, according to Ref.~\cite{Komasa2011}, or 0.04119(2)~cm$^{-1}$ as calculated with the H2Spectre code of Czachorowski~\textit{et~al.}~\cite{H2Spectre} and Komasa~\textit{et~al.}~\cite{Komasa2019}. According to Ref.~\cite{Komasa2011}, this state does not exist in the Born-Oppenheimer approximation and the corresponding wavefunction cannot be obtained using the potential energy curve of Schwenke~\cite{Schwenke1988}. The accuracy of the calculations was estimated by comparing the obtained dissociation energies with those calculated with the H2Spectre code \cite{Czachorowski2018,Komasa2019}. The average value of the relative differences is approximately 0.32\% for H$_{2}$ and 0.06\% for D$_{2}$ and is mostly attributed to weakly bound states with large values of $\nu$.

\subsection{Nuclear spin-rotation coupling constant}
\label{Sec:NSR-constant}
Nuclear spin-rotation constants were calculated at the coupled cluster with single and double (CCSD) excitation level using gauge-including atomic orbitals (GIAOs) and the uncontracted cc-pV6Z basis set~\cite{woon:94a,peterson:94a} as described in Ref.~\cite{Jozwiak2020}. The calculations were
performed for the interatomic $R$ distances in the
0.30--4.00~{\AA} range with steps of 0.01~{\AA} (for distances larger than 4.00~{\AA} the values of the nuclear spin-rotation constants drop to zero). The numerical results of the $R$-dependent coupling constants
are provided in Supplementary Materials~\cite{SupMat}. All the calculations have been performed with the CFOUR quantum-chemical package~\cite{cfour} (version 2.1).

 \subsection{Spin-spin dipole interaction}
\label{SpinSpinConstant}

 The spin-spin dipole coupling constants, Eq.~\eqref{eq:dipdip-couplingconstant}, were obtained using DVR-wavefunctions $\chi_{\nu N}$ and the fundamental constants taken from CODATA~\cite{CODATA}. 
 
 \subsection{Electric quadrupole interaction}
\label{Sec:QuadConstant}
Following the approach from Ref.~\cite{Jozwiak2020}, we employed the Born-Oppenheimer EFG reported by Pavanello~\textit{et~al.}~\cite{Pavanello2010}. 
First, we calculated the expectation value of the $R$-dependent EFG in the $v = 0$, $N = 1$ state of D$_{2}$, using the experimental quadrupole coupling constant reported by Code and Ramsey~\cite{Code1971} ($c_{\rm Q} = 225.044(24)$~kHz), and determined the value of the electric quadrupole moment, $Q_{\rm D}$:
\begin{equation}
    Q_{\rm D} = 0.28598(3)\,\,\mathrm{fm}^{2},
\end{equation}
where the uncertainty, following Refs.~\cite{Jozwiak2020} and \cite{Pavanello2010}, is propagated from the experimental standard uncertainty. We note that this value is in perfect agreement with the one reported by us in the previous work~\cite{Jozwiak2020}, based on the coupling constant for the HD molecule ($Q_{\rm D} = 0.28591(8)\,\,\mathrm{fm}^{2}$). 

\subsection{Comparison with previous results}
\label{Sec:Comparison}
 \begin{table*}[!ht]
\caption{\label{tab:tableCompare}Comparison of the calculated coupling constants for the H$_{2}$ and D$_{2}$ molecules with the available literature data}\centering
\begin{tabular}{cccccc} \hline
Isotopologue & ($\nu,N$) & $c_{\rm nsr}$ (kHz) &  $c_{\rm dip}$ (kHz) & $c_{\rm Q}$ (kHz) &\\ \hline
\multirow{9}{*}{H$_{2}$}& \multirow{3}{*}{(0,1)}  & 114.16 &  288.220  &\multirow{3}{*}{--} & \\
& & 114.299$ ^{\rm a}$ & 288.365$ ^{\rm b}$ & & Theoretical \\ 
 & & 113.904(30)$ ^{\rm c}$& 288.355(120)$ ^{\rm c}$  & & Experimental \\ \cline{2-6}
 & \multirow{3}{*}{(0,3)} & 110.89 & 281.664  & \multirow{3}{*}{--}& \\
 & & 111.021$ ^{\rm a}$ & 281.505$ ^{\rm d}$  & & Theoretical \\
  & & 111.10(25)$ ^{\rm c}$ & 281.0(1.1)$ ^{\rm c}$  & & Experimental \\ \cline{2-6}
  & \multirow{3}{*}{(0,5)} & 105.34 & 270.491  & \multirow{3}{*}{--}&  \\
 & & 105.471$ ^{\rm a}$ & 269.39$ ^{\rm d}$ & & Theoretical\\
  & & 105.37(32)$ ^{\rm d}$ & 271.2(2.3)$ ^{\rm d}$  & & Experimental \\\hline
  \multirow{6}{*}{D$_{2}$}& \multirow{3}{*}{(0,1)}  & 8.81 &  6.841 & 225.044 & \\
  &  & 8.82$ ^{\rm a}$ &  6.843$ ^{\rm b}$ & -- & Theoretical \\
  &  & 8.768(3)$ ^{\rm b}$ & --  & 225.044(24)$ ^{\rm b}$ & Experimental \\ \cline{2-6}
   & \multirow{3}{*}{(0,2)}  & 8.76 &  6.810 & 223.397 & \\
  &  & 8.769$ ^{\rm a}$ &  6.810$ ^{\rm b}$ & -- & Theoretical \\
  &  & 8.723(20)$ ^{\rm b}$ &  6.813(35)$ ^{\rm b}$ & 223.380(180)$ ^{\rm b}$ & Experimental \\\hline
\end{tabular}\\
\footnotesize{$^{\rm a}$Ref.~\cite{gauss:97a}, 
$ ^{\rm b}$Ref.~\cite{Code1971}}, $ ^{\rm c}$Ref.~\cite{Harrick1953},
$ ^{\rm d}$Ref.~\cite{Verberne1978} \\
\end{table*}

Table~\ref{tab:tableCompare} presents a comparison of our vibrationally-averaged coupling constants with the available literature data. Most of the experimental studies of H$_{2}$ and D$_{2}$, which employed molecular beam magnetic resonance method, were focused on the ${\nu=0, N=1}$ state. Values of the hyperfine coupling constants were first extracted from the radiofrequency spectra in strong magnetic field by Kellogg~\textit{et~al.}~\cite{Kellog1939,Kellog1939b} and were later refined by Kolsky~\textit{et~al.}~\cite{Kolsky1950a,Kolsky1950b} in the weak field regime. Those values were used by Ramsey~\cite{Ramsey1952a} in the theoretical analysis of the hyperfine structure of the first rotational state in both hydrogen and deuterium molecules. Up to this date, the most accurate values of the hyperfine coupling constant in this state of H$_{2}$ come from the paper of Harrick~\textit{et~al.}~\cite{Harrick1953} and we compare our results with this dataset. Hyperfine coupling constants for two more (${\nu=0, N=3}$ and ${\nu = 0, N=5}$) rotational states in H$_{2}$ were extracted by Verberne~\textit{et~al.}~\cite{Verberne1978} from experiments involving the molecular beam magnetic resonance technique in the low-field regime. In the case of the D$_{2}$ molecule, the most accurate experimental data were reported by Code and Ramsey~\cite{Code1971}, who refined the data for the ${\nu=0, N=1}$ state and performed additional measurements for the ${\nu = 0, N=2}$ rotational level.

In the case of theoretical investigations of the hyperfine coupling constants in H$_{2}$ and D$_{2}$, we compare our results with the spin-rotation constants published by Gauss~\textit{et~al.}~\cite{gauss:97a}, who reported values of $c_{\rm nsr}$ for a wide range of rovibrational levels in both the H$_{2}$ and D$_{2}$ molecules. 

Theoretical values of the spin-spin dipole coupling constant were reported for the ${\nu=0, N=1}$ (Code~\textit{et~al.}~\cite{Code1971}), ${\nu=0, N=3}$ and ${\nu=0, N=5}$ (Verberne~\textit{et~al.}~\cite{Verberne1978}) levels in H$_{2}$ and for the ${\nu=0, N=1}$ and ${\nu = 0, N=2}$ states in D$_{2}$ (Code~\textit{et~al.}~\cite{Code1971}).

The experimental value of the quadrupole coupling constant was mostly used as a reference value in theoretical determination of the deuteron's quadrupole moment from the calculations of EFG~\cite{Reid1973,Reid1975,Bishop1979,Pavanello2010} and, thus, no theoretical predictions of $c_{\rm Q}$ are available.

Following our previous work~\cite{Jozwiak2020}, we estimate the uncertainty of the nuclear-spin rotation constants at 300~Hz. Our results agree very well with the experimental data for both H$_{2}$ and D$_{2}$. We note that for the $N=3$ and $N=5$ levels in H$_{2}$, the values of $c_{\rm nsr}$ reported here lie within the experimental error bars. In the case of the spin-spin dipole constants, the uncertainty is determined by the quality of the rovibrational wavefunctions, which was estimated at the beginning of this section. In all considered cases, our values are in perfect agreement with the experimental results. In the case of the quadrupole coupling constants, the comparison concerns only the ${\nu=0, N=2}$ state, and our result lies within the experimental error bars.

\section{Determination of the energy levels}
\label{Sec:EnergyLevels}
Since neither the hyperfine Hamiltonian of H$_{2}$, Eq.~\eqref{eq:hyperfineHamiltonian}, nor the hyperfine Hamiltonian of D$_{2}$, \eqref{eq:hyperfineHamiltonianD2}, are diagonal in the coupled representation, $\ket{\nu; N I F m_{\rm F}}$, the hyperfine energy levels should be determined by their diagonalization. Owing to the fact that none of the components of Eqs.~\eqref{eq:hyperfineHamiltonian} and \eqref{eq:hyperfineHamiltonianD2} mix different $F$ and $F'$ (and $m_{\rm F}$ and $m_{\rm F}'$) states, one can diagonalize each $F$-labelled submatrix independently. In the case of \textit{ortho}-H$_{2}$ and \textit{para}-D$_{2}$, each submatrix is of the ${3 \times 3}$ size, where the columns (or rows) correspond to $N=F-1$, $N=F$ or $N=F+1$ coupled basis vectors. Due to the fact that \textit{ortho}-H$_{2}$ and \textit{para}-D$_{2}$ involve only odd rotational levels, the submatrices are of $1 \times 1$ ($N=F$) or $2 \times 2$ ($N=F-1$ and $N=F+2$) dimensions for odd (and $F=0$) and even values of $F$, respectively. In the latter case, the off-diagonal matrix elements, which couple $N=F-1$ and $N=F+1$ states, are from 9 to 11 orders of magnitude smaller than diagonal terms. Thus, in the case of \textit{ortho}-H$_{2}$ and \textit{para}-D$_{2}$, $N$ can be treated as a good quantum number and we retain the coupled basis vectors labeling for the hyperfine energy levels.

In the case of \textit{para}-D$_{2}$ ($I=0$ or 2), the submatrices are of ${6 \times 6}$ size. Due to the fact that \textit{para}-D$_{2}$ involves only even rotational quantum numbers, the dimensions of submatrices are reduced to ${2 \times 2}$ and ${4 \times 4}$, for even and odd $F$ values, respectively. The hyperfine energy levels were determined by numerical diagonalization of each submatrix. The eigenvectors of the hyperfine matrix are referred to as $\ket{\nu; N F m_{\rm F} (\pm) }$. Due to a very weak coupling between  rotational levels which differ by $\Delta N = \pm 2$, $N$ remains a good quantum number. Following the work of Code and Ramsey~\cite{Code1971}, we introduce the $(\pm)$ labels to distinguish between the eigenstates which originate from the mixing of $I=0$ and $I=2$ states and correspond to higher $(+)$ or lower $(-)$ energy. 

In the subsequent discussion, we shall employ the following relation between the coupled basis vectors and the eigenvectors:
\begin{align}
\begin{split}
    \label{eq:diagonalization}
    \ket{\nu; N F m_{\rm F} (\pm)} 
    =\sum_{I=0,2} \sum_{N' = F-I}^{F+I} a_{N' I}^{N' F (\pm) } \ket{\nu; N I F m_{F}},
\end{split}
\end{align}
where $a_{N' I}^{N F (\pm) }$ denote the mixing coefficients, obtained from the diagonalization of the Hamiltonian of \textit{ortho}-D$_{2}$.

\section{Line intensities}
\label{Sec:LineIntensities}

In this section we provide a complete list of intensities of the hyperfine components of all quadrupole lines in both the H$_{2}$ and D$_{2}$ molecules. The intensity of the quadrupole transition between two degenerate states (initial and final), in the SI units, is given~\cite{Campargue2012,Kassi2012,Karl_1967} as:
\begin{equation}
    \label{eq:lineintensity}
    S_{\rm fi} = \frac{2\pi^{4}}{15hc^{3}\epsilon_{0}}\nu_{0}^{3}C_{N_{i}}P_{\rm fi}(T)|\mathcal{Q}_{\rm fi}|^{2},
\end{equation}
where the transition frequency and the electric quadrupole transition moment are denoted by $\nu_{0}$ and $\mathcal{Q}_{\rm fi}$, respectively. $h$, $c$ and $\epsilon_{0}$ are the Planck's constant, the speed of light in vacuum and the vacuum permittivity, respectively. $C_{N_{i}}$ is an algebraic coefficient, which depends on the considered rovibrational branch:
\begin{enumerate}
    \item O branch ($N_{f} = N_{i} - 2$):
    \begin{equation}
        \label{eq:cjObranch}
        C_{N_{i}} = \frac{3N_{i}(N_{i}-1)}{2(2N_{i}+1)(2N_{i}+3)},
    \end{equation}
    \item Q branch ($N_{f} = N_{i}$):
    \begin{equation}
        \label{eq:cjQbranch}
        C_{N_{i}} = \frac{N_{i}(N_{i}+1)}{2(2N_{i}-1)(2N_{i}+3)},
    \end{equation}
    \item S branch ($N_{f} = N_{i} + 2$):
    \begin{equation}
        \label{eq:cjSbranch}
        C_{N_{i}} = \frac{3(N_{i}+1)(N_{i}+2)}{2(2N_{i}+1)(2N_{i}+3)}.
    \end{equation}
\end{enumerate}
The temperature-dependent part, $P_{\rm fi} (T)$, is given as
\begin{equation}
\label{eq:lineintensity-2}
    P_{\rm fi}(T) = w_{i} (2N_{i}+1)\frac{ \left(e^{-E_{i}/k_{\rm B}T}-e^{-E_{f}/k_{\rm B}T}\right)}{Q(T)},
\end{equation}
with the partition function, $Q(T)$, defined as:
\begin{equation}
\label{eq:lineintensity-3}
Q(T) = \sum_{k} w_{k} (2N_{k}+1) e^{-{E_{k}/k_{\rm B}T}}.
\end{equation}
{{$w_{k}$} denotes the degeneracy factor due to nuclear spin statistics. In the case of the H$_{2}$ molecule, $w_{k}$ corresponds to 1 or 3, for even and odd rotational levels, respectively. In the case of D$_{2}$, $w_{k}$ equals 3 or 6, for odd and even rotational quantum numbers, respectively. $E_{k}$ corresponds to the energy of the $k$-th rovibrational state, $k_{\rm B}$ denotes the Boltzmann constant and $T$ is the temperature.}

Following our previous work~\cite{Jozwiak2020}, apart from the line intensity defined in Eq.~\eqref{eq:lineintensity}, we provide the temperature-independent line intensity, $S_{\rm fi}/P_{\rm fi}(T)$. This parameter corresponds to the case in which the entire population of a given molecule occupies the $i$-th rovibrational state. This parameter can be used to determine line intensities at any~$T$.

Line intensity is determined by the quadrupole transition moment, $\mathcal{Q}_{\rm fi}$, which is obtained as:
\begin{equation}
    \label{eq:quadmoment-me}
    \mathcal{Q}_{\rm fi} = \int {\rm d}R \chi_{f}^{*}(R) \mathcal{Q}(R) \chi_{i}(R)
\end{equation}
$\mathcal{Q}(R)$ is the quadrupole moment function, defined~\cite{Campargue2012,Kassi2012,Karl_1967} as:
\begin{equation}
    \label{eq:quadmoment-func}
     \mathcal{Q}(R) = \frac{R^{2}}{2} - \frac{1}{2}\braket{\phi|\sum_{i}3z_{i}^{2}-r_{i}^{2}|\phi},
\end{equation}
where $z_{i}$ and $r_{i}$ refer to the coordinates of the $i$-th electron and $\ket{\phi}$ denotes the electronic wavefunction, which is parametrically dependent on the internuclear coordinate,~$R$.

Electric quadrupole moment for the H$_2$ molecule was calculated at the CCSD level with the uncontracted double-augmented of six-$\zeta$ quality (d-aug-cc-pV6Z) basis set~\cite{woon:94a,peterson:94a}. The calculations were performed for the interatomic $r_{\rm HH}$ distances in the 0.30--10.60~{\AA} range with steps of 0.01~{\AA}. The numerical results of the $R$-dependent quadrupole moment are provided in Supplementary Materials~\cite{SupMat}.

The calculations have been performed with the CFOUR~\cite{cfour} program (version 2.1).

To estimate the convergence of the calculated values of the quadrupole moment, we have evaluated the complete-basis (CBS) limit values with the two-parameter formula proposed by Helgaker \textit{et al.}~\cite{helgaker:97a}:
\begin{equation}
c(X)=c_{\rm CBS}+bX^{-3}
\end{equation}
where $c_{\rm CBS}$ and $b$ are fitted parameters and $X$ is the cardinal number of an aug-cc-pVXZ basis set. For the vibrationally-averaged geometry ($\langle$\textit{r}$_{\rm HH}\rangle =$ 0.7666393~{\AA}~\cite{roy:87a,jankowski:2005a}), the difference between the quadrupole moment in the CBS limit and calculated with the uncontracted d-aug-cc-pV6Z basis is less than $7\cdot 10^{-4}$~a.u.

Quadrupole transition moments, $\mathcal{Q}_{\rm fi}$, were evaluated for all the considered rovibrational transitions, using previously described rovibrational wavefunctions. Our results are in excellent agreement (the largest relative difference is at the level of 0.1\%) with values reported by Campargue~\textit{et~al.}~\cite{Campargue2012} and Kassi~\textit{et~al.}~\cite{Kassi2012} for H$_{2}$ and D$_{2}$ molecules, respectively.

\section{Line intensities of hyperfine transitions}
\label{Sec:HFLineIntensities}
\begin{table*}
\caption{\label{tab:tableLineListH2}Example of the calculated positions and intensities of the hyperfine components of the  quadrupole transitions in the H$_{2}$ molecule, which are provided in Supplementary Materials~\cite{SupMat}. Frequencies of the rovibrational transition are calculated with the \textrm{H2Spectre} code of Czachorowski~\textit{et~al.}~\cite{H2Spectre} and Komasa~\textit{et~al.}~\cite{Komasa2019}. Note that for the rovibrational transition, $\mathcal{Q}$ corresponds to $\mathcal{Q}_{\rm fi}$ from Eq.~\eqref{eq:lineintensity}, while for the hyperfine components, $\mathcal{Q}$ denotes $\mathcal{Q}_{\rm fi}^{\rm HF}$ from Eq.~\eqref{eq:HFlineintensity}. Intensity (in the sixth column) corresponds to the temperature-independent line intensity, $S_{\rm fi}/P_{\rm fi}(T)$ (see Sec.~\ref{Sec:LineIntensities}).}
\begin{tabular}{cccrccc}\hline
 Band& Line &Hyperfine transition  & Frequency (MHz) & $\mathcal{Q}$ (a.u.) & Intensity & Intensity at 296~K \\
  & & $\ket{N' F'} \leftarrow \ket{N F}$ & &  &(cm/molecule) & (cm/molecule)
 \\\hline
                           1-0 &                          Q(1) & &              124571373.8(6.9) &                       0.08782 &   3.29002703$\times 10^{-26}$ &  2.17001465$\times 10^{-26}$ \\
 & &$\ket{1\,2}\,\leftarrow\,\ket{1\,0}$ &                      -0.60098 &                       0.05554 &   3.65558559$\times 10^{-27}$ &  2.41112739$\times 10^{-27}$ \\
 & &$\ket{1\,2}\,\leftarrow\,\ket{1\,1}$ &                      -0.05448 &                       0.08332 &   8.22506758$\times 10^{-27}$ &  5.42503662$\times 10^{-27}$ \\
 & &$\ket{1\,2}\,\leftarrow\,\ket{1\,2}$ &                       0.00091 &                       0.07348 &   6.39727479$\times 10^{-27}$ &  4.21947293$\times 10^{-27}$ \\
 & &$\ket{1\,1}\,\leftarrow\,\ket{1\,1}$ &                       0.00323 &                       0.04810 &   2.74168919$\times 10^{-27}$ &  1.80834554$\times 10^{-27}$ \\
 & &$\ket{1\,1}\,\leftarrow\,\ket{1\,2}$ &                       0.05862 &                       0.08332 &   8.22506758$\times 10^{-27}$ &  5.42503662$\times 10^{-27}$ \\
 & &$\ket{1\,0}\,\leftarrow\,\ket{1\,2}$ &                       0.58765 &                       0.05554 &   3.65558559$\times 10^{-27}$ &  2.41112739$\times 10^{-27}$ \\
 \hline
\end{tabular}
\end{table*}
\begin{table*}
\caption{\label{tab:tableLineListD2}Example of the calculated positions and intensities of the hyperfine components of the  quadrupole transitions in the D$_{2}$ molecule, which are provided in Supplementary Materials~\cite{SupMat}. Frequencies of the rovibrational transition are calculated with the \textrm{H2Spectre} code of Czachorowski~\textit{et~al.}~\cite{H2Spectre} and Komasa~\textit{et~al.}~\cite{Komasa2019}. Note that for the rovibrational transition, $\mathcal{Q}$ corresponds to $\mathcal{Q}_{\rm fi}$ from Eq.~\eqref{eq:lineintensity}, while for the hyperfine components, $\mathcal{Q}$ denotes $\mathcal{Q}_{\rm fi}^{\rm HF}$ from Eq.~\eqref{eq:HFlineintensity}. Intensity (in the sixth column) corresponds to the temperature-independent line intensity, $S_{\rm fi}/P_{\rm fi}(T)$ (see Sec.~\ref{Sec:LineIntensities}).}
\begin{tabular}{cccrccc}\hline
 Band& Line &Hyperfine transition  & Frequency (MHz) & $\mathcal{Q}$ (a.u.) & Intensity & Intensity at 296~K \\
  & & $\ket{N' F'\pm} \leftarrow \ket{N F\pm}$ & &  &(cm/molecule) & (cm/molecule)
 \\\hline
                            2-0 &                          S(2) & &              187104298.7(4.5) &                       0.00818 &   1.24370481$\times 10^{-27}$ &  4.78449686$\times 10^{-28}$ \\
 & &$\ket{4\,4-}\,\leftarrow\,\ket{2\,2+}$ &                      -0.17198 &                       0.00225 &   6.09118228$\times 10^{-30}$ &  2.34326042$\times 10^{-30}$ \\
 & &$\ket{4\,4-}\,\leftarrow\,\ket{2\,3}$ &                      -0.11172 &                       0.00462 &   2.57010972$\times 10^{-29}$ &  9.88713868$\times 10^{-30}$ \\
 & &$\ket{4\,2}\,\leftarrow\,\ket{2\,2+}$ &                      -0.10029 &                       0.00297 &   1.06198195$\times 10^{-29}$ &  4.08541421$\times 10^{-30}$ \\
 & &$\ket{4\,3}\,\leftarrow\,\ket{2\,2+}$ &                      -0.06451 &                       0.00621 &   4.64617102$\times 10^{-29}$ &  1.78736872$\times 10^{-29}$ \\
 & &$\ket{4\,2}\,\leftarrow\,\ket{2\,3}$ &                      -0.04003 &                       0.00111 &   1.48060096$\times 10^{-30}$ &  5.69582959$\times 10^{-31}$ \\
 & &$\ket{4\,6}\,\leftarrow\,\ket{2\,4}$ &                      -0.03829 &                       0.01577 &   2.99410416$\times 10^{-28}$ &  1.15182332$\times 10^{-28}$ \\
 & &$\ket{4\,5}\,\leftarrow\,\ket{2\,3}$ &                      -0.02974 &                       0.01256 &   1.90010457$\times 10^{-28}$ &  7.30964797$\times 10^{-29}$ \\
 & &$\ket{4\,3}\,\leftarrow\,\ket{2\,3}$ &                      -0.00426 &                       0.00379 &   1.72736779$\times 10^{-29}$ &  6.64513452$\times 10^{-30}$ \\
 & &$\ket{4\,4+}\,\leftarrow\,\ket{2\,2+}$ &                       0.00444 &                       0.01094 &   1.44111423$\times 10^{-28}$ &  5.54392525$\times 10^{-29}$ \\
 & &$\ket{4\,4-}\,\leftarrow\,\ket{2\,4}$ &                       0.00738 &                       0.00158 &   3.00402435$\times 10^{-30}$ &  1.15563959$\times 10^{-30}$ \\
 & &$\ket{4\,4-}\,\leftarrow\,\ket{2\,2-}$ &                       0.01333 &                       0.01197 &   1.72487831$\times 10^{-28}$ &  6.63555756$\times 10^{-29}$ \\
 & &$\ket{4\,2}\,\leftarrow\,\ket{2\,1}$ &                       0.02126 &                       0.00677 &   5.52757692$\times 10^{-29}$ &  2.12644305$\times 10^{-29}$ \\
 & &$\ket{4\,3}\,\leftarrow\,\ket{2\,1}$ &                       0.05703 &                       0.00757 &   6.90947115$\times 10^{-29}$ &  2.65805381$\times 10^{-29}$ \\
 & &$\ket{4\,2}\,\leftarrow\,\ket{2\,0}$ &                       0.06154 &                       0.00587 &   4.14568269$\times 10^{-29}$ &  1.59483229$\times 10^{-29}$ \\
 & &$\ket{4\,4+}\,\leftarrow\,\ket{2\,3}$ &                       0.06469 &                       0.00680 &   5.57319556$\times 10^{-29}$ &  2.14399241$\times 10^{-29}$ \\
 & &$\ket{4\,2}\,\leftarrow\,\ket{2\,4}$ &                       0.07907 &                       0.00014 &   2.35016026$\times 10^{-32}$ &  9.04099935$\times 10^{-33}$ \\
 & &$\ket{4\,2}\,\leftarrow\,\ket{2\,2-}$ &                       0.08502 &                       0.00229 &   6.30133434$\times 10^{-30}$ &  2.42410532$\times 10^{-30}$ \\
 & &$\ket{4\,5}\,\leftarrow\,\ket{2\,4}$ &                       0.08937 &                       0.00725 &   6.33368189$\times 10^{-29}$ &  2.43654932$\times 10^{-29}$ \\
 & &$\ket{4\,3}\,\leftarrow\,\ket{2\,4}$ &                       0.11484 &                       0.00083 &   8.22556089$\times 10^{-31}$ &  3.16434977$\times 10^{-31}$ \\
 & &$\ket{4\,3}\,\leftarrow\,\ket{2\,2-}$ &                       0.12079 &                       0.00478 &   2.75683378$\times 10^{-29}$ &  1.06054608$\times 10^{-29}$ \\
 & &$\ket{4\,4+}\,\leftarrow\,\ket{2\,4}$ &                       0.18379 &                       0.00233 &   6.51412468$\times 10^{-30}$ &  2.50596515$\times 10^{-30}$ \\
 & &$\ket{4\,4+}\,\leftarrow\,\ket{2\,2-}$ &                       0.18974 &                       0.00088 &   9.26631702$\times 10^{-31}$ &  3.56472568$\times 10^{-31}$ \\
 \hline
\end{tabular}
\end{table*}
Similarly to Eq.~\eqref{eq:lineintensity}, one can calculate the intensity of the hyperfine components of each rovibrational line:
\begin{equation}
    \label{eq:HFlineintensity}
    S^{\rm HF}_{\rm fi} = \frac{2\pi^{4}}{15hc^{3}\epsilon_{0}}\nu_{0}^{3}\frac{1}{w_{i}\left(2N_{i}+1\right)}P_{\rm fi}(T)\left|\mathcal{Q}^{\rm HF}_{\rm fi}\right|^{2},
\end{equation}
where $\mathcal{Q}^{\rm HF}_{\rm fi}$ is the reduced matrix element of the rank-2 quadrupole moment tensor, which, for \textit{ortho}-H$_{2}$ and \textit{para}-D$_{2}$ is simply:
\begin{equation}
\left|\mathcal{Q}^{\rm HF}_{\rm fi}\right|^{2} = \left|\braket{\nu_{f}; N_{f} I_{f} F_{f} m_{{\rm F}_{f}}\|\mathrm{T}^{(2)}(\mathbf{\mathcal{Q}})\|\nu_{i} ; N_{i} I_{i} F_{i} m_{{\rm F}_{i}}}\right|^{2} ,    
\end{equation}
while for \textit{ortho}-D$_{2}$ corresponds to:
\begin{align}
\begin{split}
    \label{eq:HFdipolemoment}
    &\left|\mathcal{Q}^{\rm HF}_{\rm fi}\right|^{2} =\left|\braket{\nu_{f};N_{f}F_{f}m_{{\rm F}_{f}}(\pm)\|\mathrm{T}^{(2)}(\mathbf{\mathcal{Q}})\|\nu_{i};N_{i}F_{i}m_{{\rm F}_{i}}(\pm)}\right|^{2} \\
    &= \Bigl|\sum_{I_{f}=0,2} \sum_{N_{f}' = F_{{f}}-I_{f}}^{F_{{f}}+I_{f}}
    \sum_{I_{i}=0,2} \sum_{N_{i}' = F_{{i}}-I_{i}}^{F_{{i}}+I_{i}} 
    a_{N_{f}' I_{f}}^{N_{f} F_{f} (\pm) } a_{N_{i}' I_{i}}^{N_{i} F_{i} (\pm)} \\ &\times\braket{\nu_{f}; N_{f}' I_{f} F_{f} m_{{\rm F}_{f}}\|\mathrm{T}^{(2)}(\mathbf{\mathcal{Q}})\|\nu_{i} ; N_{i}' I_{i} F_{i} m_{{\rm F}_{i}}}\Bigr|^{2} .
\end{split}
\end{align}
The last term is evaluated using the spherical tensor algebra – $\mathrm{T}^{(2)}(\mathbf{\mathcal{Q}})$ is a spherical tensor of rank-2, which acts only one part of a coupled scheme, causing a possible change of rotational quantum number, N. The reduced matrix element is evaluated as (see Chapter~5.5.4.~of~Ref.~\cite{BrownCarrington2003}):
\begin{align}
\begin{split}
    \label{eq:HFdipolemoment-2}
    &\braket{\nu_{f}; N_{f} I_{f} F_{f} m_{{\rm F}_{f}}\|\mathrm{T}^{(2)}(\mathbf{\mathcal{Q}})\|\nu_{i}; N_{i} I_{i} F_{i} m_{{\rm F}_{i}}}
    \\
    &=\delta_{I_{i}I_{f}}(-1)^{N_{f}+I_{f}+F_{i}} \sqrt{(2F_{i}+1)(2F_{f}+1)} \\
    &\times
    \begin{Bmatrix}
    N_{f} & N_{i} & 2 \\
    F_{i} & F_{f} & I_{i}
    \end{Bmatrix}
    \braket{\nu_{f}N_{f}\|\mathrm{T}^{(2)}(\mathbf{\mathcal{Q}})\|\nu_{i}N_{i}}.
\end{split}
\end{align}
Triangle conditions, imposed on the elements of the 6-\textit{j} symbol, lead to the selection rule for the total angular momentum, namely $\Delta F = F_{f}-F_{i} = 0, \pm 1, \pm 2$ (0$\nleftrightarrow$0, i.e. the transitions between ${F_{i} = 0}$ and ${F_{f} = 0}$ are forbidden).

Finally, $\braket{\nu_{f}N_{f}\|\mathrm{T}^{(2)}(\mathbf{\mathcal{Q}})\|\nu_{i}N_{i}}$ is related (in the~$X^{1}\Sigma^{+}_{g}$ state) to the transition moment defined in Eq.~\eqref{eq:quadmoment-me} in the following way:
\begin{align}
\begin{split}
    \label{eq:HFdipolemoment-3}
   &\braket{\nu_{f}N_{f}\|\mathrm{T}^{(2)}(\mathbf{\mathcal{Q}})\|\nu_{i}N_{i}} \\
   &= (-1)^{N_{f}} \sqrt{(2N_{f}+1)(2N_{i}+1)} 
   \begin{pmatrix}
       N_{f} & 2 & N_{i} \\
       0 & 0 & 0
   \end{pmatrix} \mathcal{Q}_{\rm fi}.
\end{split}
\end{align}
Evaluation of the 3-\textit{j} symbol in the last equation using tabulated values (see, e.g. Appendix C of Ref.~\cite{BrownCarrington2003}), leads to the conclusion that the squared value of the algebraic factor on the right side of Eq.~\eqref{eq:HFdipolemoment-3} corresponds to the $C_{N_{i}}$ factor (Eqs.~\eqref{eq:cjObranch}-\eqref{eq:cjSbranch}), which appears in Eq.~\eqref{eq:lineintensity} but is absent in Eq.~\eqref{eq:HFlineintensity}.

\section{Example of the complete dataset record}
\label{Sec:example}
We used the obtained hyperfine coupling constants and transition moments to calculate positions and intensities of the hyperfine components of all the rovibrational quadrupole transitions in the ground electronic states in H$_{2}$ and D$_{2}$. Our calculations result in total 220~997 hyperfine components of 16~079 quadrupole lines from the O, Q and S branches. The complete dataset can be found in Supplementary Materials~\cite{SupMat} for this article. We report line intensities at the reference temperature of 296~K (following the convention adopted in the HITRAN database \cite{HITRAN}) as well as the temperature-independent line intensities, $S_{\rm fi}/P_{\rm fi}$, from which one can calculate the spectral line intensity at any $T$. We remind the reader, that in the case of \textit{ortho}-H$_{2}$ and \textit{para}-D$_{2}$ ($I=1$) we use the $\ket{N F}$ labels to denote each hyperfine energy level and in the case of \textit{ortho}-D$_{2}$ we label the hyperfine levels as $\ket{N F \pm}$, due to the mixing of $I=0$ and $I=2$ states (see Sec.~\ref{Sec:EnergyLevels}).

Here, we present one example for each isotopologue, namely the Q(1) line from the fundamental band in H$_{2}$ and the S(2) line from the first overtone in the D$_{2}$ molecule. The Q(1) line from the fundamental band of H$_{2}$ is currently a subject of the experimental investigation using the comb-calibrated stimulated Raman spectroscopy~\cite{Lamperti:19}, which aims at sub-MHz accuracy of the unperturbed line position. This result will improve the current state-of-the-art accuracy~\cite{Dickenson2013} by a factor of at least five. Table~\ref{tab:tableLineListH2} gathers all the six hyperfine components of this rovibrational transition, which are distributed (almost symmetrically) in the range of about 1.2~MHz around the central frequency. 

The 2-0 S(2) line in D$_{2}$ has been a subject of interest of various experimental investigations~\cite{Mondelain2016,Wcislo2018,Zaborowski2020}. Currently, the most accurate measurement of this quadrupole line reached the accuracy of 161~kHz~\cite{Zaborowski2020}, which makes it the most precise measurement of the transition frequency in the D$_{2}$ molecule. Table~\ref{tab:tableLineListD2} gathers the twenty two hyperfine components of this quadrupole transition, which are distributed about 360~kHz around the central frequency. 

\section{Conclusion}
\label{Sec:Conclusion}
We have reported a list of the hyperfine coupling constants, i.e. the spin-rotation, spin-spin and, in case of the D$_{2}$ molecule, quadrupole coupling constants, for all bound states of the H$_{2}$ and D$_{2}$ isotopologues. Our results agree very well with the experimental data of Harrick~\textit{et~al}~\cite{Harrick1953}, Code~\textit{et~al}~\cite{Code1971} and Verberne~\textit{et~al}~\cite{Verberne1978}. Our calculations of the hyperfine splittings of all bound states of the two isotopologues resulted in a comprehensive list of the positions and intensities of the 220~997 hyperfine components of the 16~079 rovibrational quadrupole transitions. The intensities, positions, as well as the coupling constants and the quadrupole transition moment function, can be found in Supplementary Materials~\cite{SupMat}. The results presented here are useful as a reference data for a proper interpretation of accurate measurements of rovibrational quadrupole lines in H$_{2}$ and D$_{2}$, which are used for testing the quantum electrodynamics for molecules and searching for new physics beyond the Standard Model.

\section*{Acknowledgements}
The research is financed from the budgetary funds on science projected for 2019--2023 as a research project under the "Diamentowy Grant" program. 
P.W. is supported by the National Science Centre in Poland, Project No. 2019/35/B/ST2/01118.
Calculations have been carried out using resources provided by Wroclaw Centre for Networking and Supercomputing (\url{http://wcss.pl}), Grant No. 294.

\bibliography{mybibfile}

\end{document}